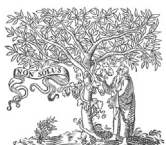
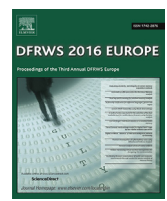

DFRWS 2016 Europe — Proceedings of the Third Annual DFRWS Europe

# Tiered forensic methodology model for Digital Field Triage by non-digital evidence specialists

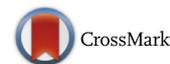

Ben Hitchcock [a, b, *], Nhien-An Le-Khac [b], Mark Scanlon [b]

[a] Royal Canadian Mounted Police, Canada
[b] School of Computer Science, University College Dublin, Ireland

## abstract

Due to budgetary constraints and the high level of training required, digital forensic analysts are in short supply in police forces the world over. This inevitably leads to a prolonged time taken between an investigator sending the digital evidence for analysis and receiving the analytical report back. In an attempt to expedite this procedure, various process models have been created to place the forensic analyst in the field conducting a triage of the digital evidence. By conducting triage in the field, an investigator is able to act upon pertinent information quicker, while waiting on the full report.

The work presented as part of this paper focuses on the training of front-line personnel in the field triage process, without the need of a forensic analyst attending the scene. The premise has been successfully implemented within regular/non-digital forensics, i.e., crime scene investigation. In that field, front-line members have been trained in specific tasks to supplement the trained specialists. The concept of front-line members conducting triage of digital evidence in the field is achieved through the development of a new process model providing guidance to these members. To prove the model's viability, an implementation of this new process model is presented and evaluated. The results outlined demonstrate how a tiered response involving digital evidence specialists and non-specialists can better deal with the increasing number of investigations involving digital evidence.

© 2016 The Authors. Published by Elsevier Ltd on behalf of DFRWS. This is an open access article under the CC BY-NC-ND license (http://creativecommons.org/licenses/by-nc-nd/4.0/).

## Introduction

In the modern policing environment there is an increasing demand to "do more with less". Within the field of Digital Forensics, this demand is manifesting itself through increasing delays for both the investigators and the court systems receiving an in-depth analytical report. The issue of the increasing backlog of digital evidence waiting to be examined in many police departments is well documented (Mislan et al., 2010; Casey et al., 2009; James and Gladyshev, 2015). A significant bottleneck during an investigation involving digital evidence is the time delay from digital evidence being sent to a specialist Technological Crime Unit (TCU) and the assignment to a forensic analyst to complete the necessary in-depth analysis and reporting. As a result, investigators are left without potentially actionable information at a time that it would be most beneficial. Society is becoming increasingly connected through the use of computers, tablets and smartphones, and this is reflected in the use of these devices by criminals. The bottleneck thus increases the backlog of files at the TCU and produces delays that have a detrimental effect on the accused's right to a speedy trial, e.g., the right "to be tried in a reasonable time" from Section 11(b) of the Canadian Charter of Rights and Freedoms.

* Corresponding author. Royal Canadian Mounted Police, Canada.
E-mail addresses: ben.hitchcock@rcmp-grc.gc.ca (B. Hitchcock), an.lekhac@ucd.ie (N.-A. Le-Khac), mark.scanlon@ucd.ie (M. Scanlon).

There are numerous methodologies that have identified a need for onsite/field analysis of the digital evidence. An informal review conducted as part of the work presented in this paper revealed backlogs ranging from one to four years James and Gladyshev (James and Gladyshev, 2015), Garda Síochána Inspectorate (Garda Síochána Inspectorate, 2015). The problem faced is that there remains limited resources dedicated to forensic analysis, including a limited number of fully trained forensic personnel. When a forensic analyst attends a search scene to provide onsite analysis, they are no longer available to conduct analysis on their assigned files in the forensic laboratory. The investigation that the forensic analyst provides the onsite/field analysis is now the highest priority file (irrespective of the severity of the crime) as it has the analysts full attention. This requirement contributes to increasing the backlog of files at the forensic laboratory.

The process model proposed as part of this paper expands on the triage component and includes a component where some forensic analyst functions are offloaded to specifically trained personnel. These personnel would be front line investigators who receive basic training in the area of forensic analysis. The work these trained personnel complete must maintain the forensic integrity of the digital evidence. These trained members need not be dedicated processors of digital evidence, but would be investigators with additional skills and training. Within the Royal Canadian Mounted Police, this practice has already been implemented by Forensic Identification Services. Certain activities, e.g., the ability to lift fingerprints from a crime scene, have been offloaded to specially trained front line personnel with great success. The investigator who lifts the fingerprint is not expected to conduct a fingerprint comparison, that is the job of the forensic identification specialist, however, the specialist now does not have to attend every scene.

The proposed end result is for a Digital Field Triage model to provide an increase in investigational efficiency and a reduction in the backlog of digital evidence waiting for analysis by a TCU. It is the aim of this paper to provide a framework whereby front-line investigators of Law Enforcement agencies are trained in the use Digital Field Triage. This framework would provide a starting point for the Digital Field Triage that could then be tailored for the type of agency by the appropriate TCU. These agencies could range from a Municipal/City Police Force to a Nationwide Law Enforcement Agency, so the trained front-line personnel could be Detectives in the same building, to Police Officers in remote locations, however, the principles of how these personnel approach the digital evidence is consistent. These personnel will also be able to contribute to an assessment of the severity of the crime and help to establish priorities.

### Contribution of this work

Currently there is no standardised approach to utilising trained non-forensic personnel in the initial stages of an investigation involving digital evidence. This is a "problem centred approach", which according to the Design Science Research Process model would make the entry point at the first activity. The work presented as part of this paper aims to provide a framework for a tiered response to digital evidence investigations where the initial stages of the investigation, i.e., those conducted outside of a forensic laboratory, are conducted by non-digital evidence specialists. More specifically, there are two primary objectives of this research:

1. Increase the efficiency of an investigation by providing artefacts from digital evidence in a timely manner.
2. Decrease the backlog of files for analysis by digital evidence specialists at a forensic laboratory.

To achieve these objectives a formal model will be developed that a non-digital evidence specialist would employ when handling digital evidence, as well as an overview of implementation.

## Background reading

### Computer Forensics Field Triage Process Model

The Computer Forensics Field Triage Model, proposed by Rogers et al. (2006), is an overall examination of the need for Field Triage to be part of any Forensic Methodology. Rogers et al. (2006). identified that individuals involved in deviant and criminal behaviour have embraced technology as a method for improving or extending their criminal trade-craft. A review was conducted of investigative models developed to assist law enforcement in the processing of digital based evidence. The various models all attempted to deal with the entire process related to the analysis of the digital evidence. This process is time consuming given the volume of digital evidence, and this still requires the digital evidence to be transferred to a central location for analysis. This process does fail when considering time critical situations such as child luring, kidnapping, and terrorist threats for example. It was determined in these situations the need for quick information and investigative leads outweighs the need for an in-depth analysis of all the potential digital evidence.

### Process model

The Computer Forensics Field Triage Process Model (CFFTPM) is defined as: *Those investigative processes that are conducted within the first few hours of an investigation, that provide information used during the suspect interview and search execution phase. Due to the need for information to be obtained in a relatively short time frame, the model usually involves an on site/field analysis of the computer system(s) in question* (Rogers et al., 2006).

The foci of the model are to:

1. Find usable evidence immediately.
2. Identify victims at acute risk.
3. Guide the ongoing investigation.
4. Identify potential charges.
5. Accurately assess the offender's danger to society.

The above aims of the model need to be adhered to while at the same time protecting the integrity of the evidence and/or potential evidence for further examination and analysis.

The strength of this model is the ability to provide the investigative team with leads and information quickly and efficiently. The approach of this process model was not to deal with the entire process, but to expand on the onsite/field aspect by identifying general and case specific phases. These six phases constitute a high level of categorization and each phase has several sub-tasks and considerations that vary according to such things as the specifics of the case, file system and operating system under investigation.

*Phases*

The initial phase of the CFFTPM identifies that proper preparation and planning is a key part of any investigation. This ranges from logistics to briefing on the crime type and other actionable intelligence. Following the planning stage is the triage itself:

- *A process in which things are ranked in terms of importance or priority.* Essentially, those items, pieces of evidence or potential containers of evidence that are the most important or the most volatile need to be dealt with first (Rogers et al., 2006).

Triage is fundamental to the CFFTPM and in conjunction with the Planning phase is what the following phases are built upon. The investigators and interviewers who are dealing directly with the suspect or witnesses need to be providing direct input to the computer forensic examiner at this stage. This ensures that correct prioritization and assumptions are being made as it is often the case that this is the first time the computer forensic examiner has been involved in the investigation and so has no case knowledge.

The following stages provide details on the types of artefacts that the forensic analyst should look for to assist in the gathering of evidence. The stages divide the artefacts into general artefacts and case specific artefacts. The general artefacts are related to usage/user profiles, chronology/timeline, and the internet and would be similar for every investigation. The case specific artefacts adjusts the focus of the forensic examiner to those artefacts specifically related to the current investigation. For example in a child pornography investigation the highest priority should be given to the graphic and audio visual files containing the child pornography. Of note though is that a "traditional" examination would likely involve a thorough examination of all of these artefacts as well as many others. The mandates of the CFFTPM require that the examiner judiciously evaluate the potential benefit of examining each of these artefacts with the time cost of doing so.

*Discussion*

The CFFTPM identified that in time sensitive investigations there is a need to obtain artefacts from digital evidence quicker and this is best done in the field. The important point is that following the CFFTPM a computer forensic examiner has not precluded a more thorough traditional examination and analysis back in the lab. Throughout the process the procedures used have maintained both the forensic integrity of the digital evidence and the chain of custody.

The CFFTPM is an excellent starting point, however, it does rely on the use of trained forensic analysts. The ideas formulated in the CFFTPM will be used as a basis for a model to utilize non-forensic analysts who will be able to complete similar tasks in the field.

*ISO 27037*

ISO 27037 is entitled "Information technology – Security Techniques – Guidelines for identification, collection, acquisition, and preservation of digital evidence" (ISO 27037, 2012). This international standard provides guidelines for specific activities and the persons responsible in handling potential digital evidence and defined their processes for potential digital evidence. Two of the defined positions were Digital Evidence First Responders (DEFR) responsible for identification and Digital Evidence Specialists (DES) responsible for collection.

A number of possible methodologies can be developed which could be certified to follow the standard. The standard describes what should be done, not how it should be done. For example: a forensic copy should be created, clearly identified in context and tracked, not that the XYZ tool should be used. Within the context of this ISO the digital evidence and actions of the DEFR and the DES must include the concepts of Audibility, Repeatability and Reproducibility.

This ISO does bring in the concept of triage by recognizing that handling of digital evidence needs to create balance between the drivers of evidential quality, timeliness of analysis, restoration of service and cost of digital evidence collection. Any prioritization related to this balance needs to minimize the risk of potential digital evidence being spoiled and maximize evidentiary value of potential digital evidence collection.

Digital evidence needs to be governed by three fundamental principles:

- Relevance – Digital evidence is relevant when it goes towards proving or disproving an element of the specific case being investigated.
- Reliable – To ensure digital evidence is what it purports to be.
- Sufficiency – The collection of enough potential digital evidence to allow the elements of the matter to be adequately examined or investigated. - Understanding this concept is important to prioritize the effort properly when time or cost is a concern.

The tiered forensic response to an investigation will be based on the role of the DEFR, however, the responsibilities will be reduced to provide investigational assistance using personnel who have other investigational responsibilities than dealing primarily with digital evidence. Though the principles of DEFR needing to display a level of competency in handling digital evidence using the tools and methods selected will be adhered to.

*Backlog*

Prior to the application of any form of triage, the forensic process was time consuming as identified in the CFFTPM (Rogers et al., 2006). During the forensic process there were natural bottlenecks, primarily as the number of investigators in the field are greater than the number of TCU analysts available. While the TCU analysts are conducting the acquisition, analysis, and report writing for one investigation, additional investigations are coming in and are placed in a queue to await assignment. As the queue grows the backlog of investigations becomes more and more evident.

With the creation of a backlog a mechanism is required to determine which of the investigations should be assigned first. A time based mechanism of first in, first out does not take into account the severity of the crime, so an identification of stolen property file would take precedence over a hands on paedophile investigation. The standard mechanism used is one based on the severity of the crime, so crimes against a person take higher priority. A large proportion of investigations involve child pornography and there is always a possibility of a hands on offender. Naturally these type of investigations receive the priority, however, this means that fraud related files are continually given lower priorities.

The CFFTPM shows the benefit of crime scene attendance, but still identifies the need for further analysis at a later date. The question then is whether the TCU analyst continues with the analysis of the digital evidence they examined at the search scene, or does this follow up analysis wait until the file has worked its way up in the backlog. Either way the backlog queue is not being reduced, and taking into account travel and preparation time this technique may even create a greater backlog.

Attempts to alleviate the backlog included hiring technicians to only receive and acquire the digital evidence, so the TCU analyst would only need to concentrate on the analysis and report writing components. In some cases, this idea did reduce the backlog queue, but nowhere did it remove it to an acceptable level, merely moved the bottleneck further along the process. The concept of triage was also introduced within the lab setting. This did reduce the time taken to complete the analysis as those exhibits without relevant artefacts did not receive a full analysis. With the increasing number of items of digital evidence seized during the course of the investigation, the TCU analyst is still having to deal with every exhibit forwarded by the investigator which again is time consuming.

**Digital Field Triage**

Digital Field Triage (DFT) is designed to provide the knowledge, skills and abilities for non-digital evidence specialists to conduct limited forensic activities (Rogers et al., 2006). For the DFT to work there are three fundamental concepts:

- DFT cannot work in isolation and must work with a parent TCU.
- DFT must maintain the forensic integrity of the digital evidence.
- A DFT assessment does not replace a TCU analysis.

To ensure a standardized approach the DFT members would receive training, continuous support and management from a parent TCU. The parent TCU is responsible for safeguarding the program through adherence to policy, reviews and continuous assessments. A DFT member is responsible for identifying, or assessing, which items of digital evidence contain artefacts related to the offence under investigation. In the example of a child pornography investigation the DFT assessment would meet the required threshold for further analysis if illicit images are located on the computer. The DFT member is able to state that they observed a number of illicit images on the computer, but they are not able to provide information related to how the images got there, the location of the images, and other pieces of information only a trained forensic analyst could provide. Fig. 1 is used to provide a simple overview of DFT members role and TCU members role.

*Digital Field Triage Model*

The proposed Digital Field Triage model follows four phases which are loosely based on the CFFTPM (Rogers et al., 2006), but designed with a DFT member in mind and not a forensic analyst (See Fig. 2).

*Planning (how will the investigation be dealt with?)*

In the initial stages of the investigation the DFT member provides assistance to the investigator as a resource person in the area of digital evidence. As the investigation progresses towards the execution of a search warrant the DFT member provides further assistance as to the specifics of the actual search. This covers a risk assessment including topics such as:

- Is it a mission-critical digital device that cannot tolerate any downtime (ISO 27037, 2012)?
- Is it within the DFT member's comfort zone?
- What are the suspect's abilities?
- What is the crime type being investigated?

As the DFT member is has been part of the investigation from the outset, then the DFT member is privy to the all important case knowledge.

*Assessment (how will the relevant artefacts be located?)*

In conjunction with the investigative team, the DFT member identifies the digital evidence at the search scene and processes it accordingly. Each item of digital evidence is prioritized as to the likelihood of it containing the relevant artefacts. A simple example would be the computer belonging to a known sex offender would be prioritized over that of a room-mate with no criminal record. Though not foolproof, it does rely on the balance of probabilities. This is similar to Social Analysis where the suspect and the complainant are profiled, but expanded to all residents of the search location (James and Gladyshev, 2013). Before any further assessment is done, the DFT member must again conduct a risk analysis, as was conducted in the planning stages. The DFT member must still operate within their comfort zone and within the scope of the tool. When the digital evidence has been prioritized the DFT member is able to conduct an assessment of the digital evidence using the TCU approved tool and methodology, as well as working within the parameters of the identified crime type.

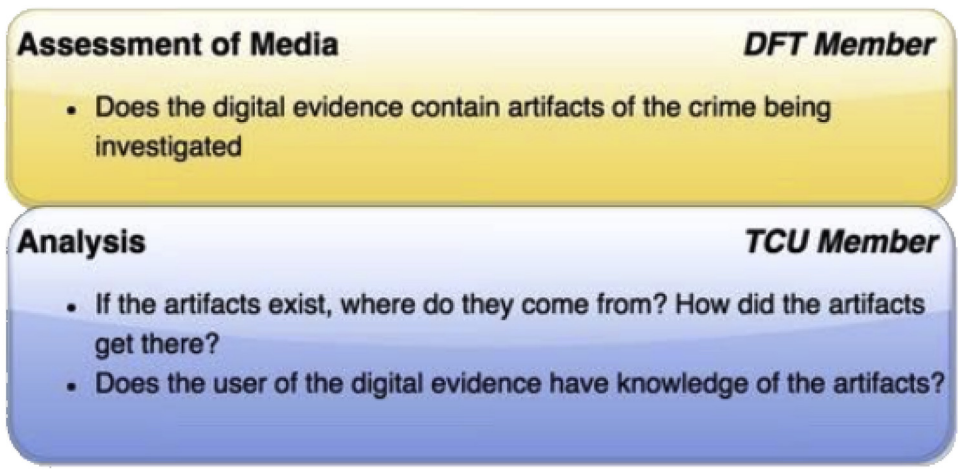

Fig. 1. Digital Field Triage Member's role.

Usually a computer would be one of the first items to be assessed, so the TCU approved tool would create a list of items that have been attached. This information would provide the investigators with a list of additional digital evidence to look for. The items that have been attached to a priority computer, would then also be given a high priority.

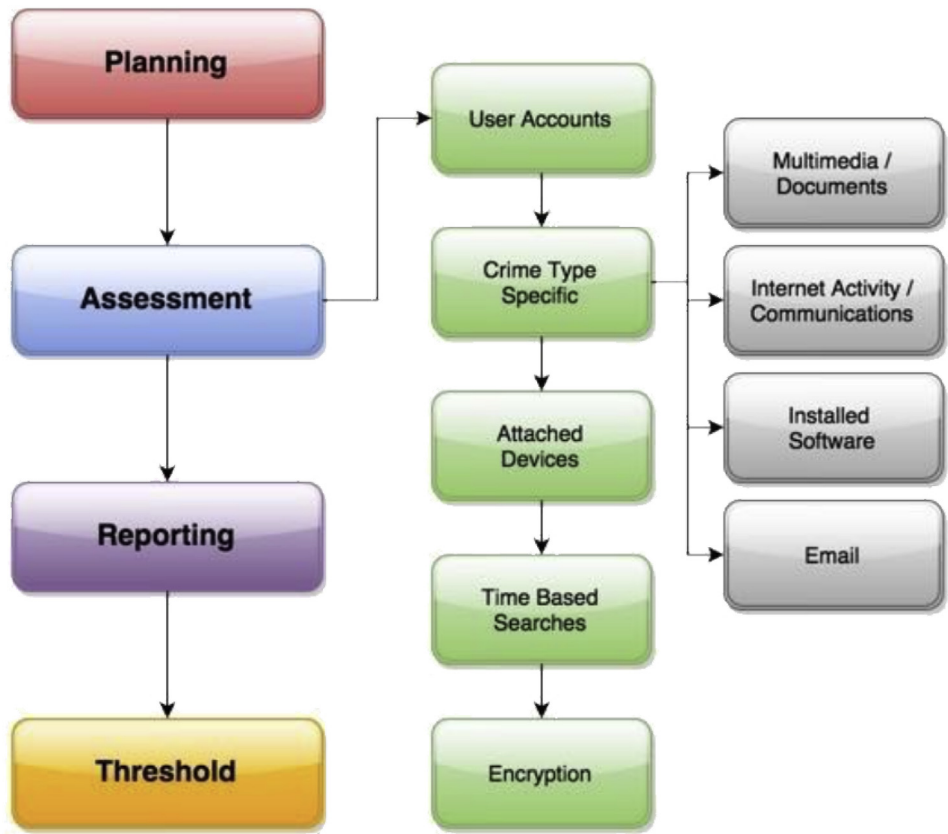

Fig. 2. Digital Field Triage model.

*Reporting (document activity and artefacts located)*

Upon completion of an assessment, a DFT member creates an *Observation Report*. Depending on the TCU approved tool the artefacts extracted may be "noisy" as a number of documents or photographs could have been located. The DFT member then looks over the raw data and brings the investigator's attention to salient artefacts. In an identity theft investigation, the fake identifications, passport style photographs, and other related artefacts are of interest and would be highlighted, but the investigator and the DFT member would still have access to review all the photographs. The *Observation Report* is simply an observation of the extracted artefacts and not an analysis report, so the DFT member will not make any opinion based statements related to the artefacts. The DFT member will also include any notes created and a listing of the searches they conducted using the TCU approved tool.

*Threshold (what is to be done with the digital evidence?)*

The DFT member and the investigator determine if the artefacts extracted and observed meet the required threshold for further TCU analysis. This phase relies on the strength of the TCU approved tool and the training of the DFT member for an accurate determination to be made.

*Benefits of DFT model*

The DFT model provides a means of increasing the operational efficiency of an investigation by providing investigators with actionable intelligence when it is most needed. This actionable intelligence may be in the form of identifying further avenues of investigation or providing artefacts that could be presented to the suspect during the initial interview.

With an assessment of the digital evidence being conducted during the initial stages of the investigation, only the items of relevance, or meeting the threshold, are forwarded to the TCU. This then cuts down the items of digital evidence being forwarded to the TCU, and so would reduce the backlog. The investigator also has the ability to share the *Observation Report* with a prosecutor and then a determination could be made that the only digital evidence required could be a specific artefact, like a document or an email. The TCU analyst need only prepare a targeted report covering that specific artefact. This then reduces the time spent on an analysis by a TCU analyst, thus increasing efficiency and reducing the backlog.

*Risks*

The DFT model does have inherent risks associated with it, and it is the management, training and TCU supported tools that need to mitigate these risks. The first risk is always excluding an item of digital evidence that is important to the investigation. One paper conducted research on this exact topic and found that an advanced preview was found to be effective at reducing the number of items of digital evidence needing to receive a full analysis while not excluding any items that contained artefacts related to the offence (James and Gladyshev, 2013).

There is also a risk in not utilizing a DFT model as TCU analyst's consider what the "sufficiency of examination" is, or in other terms considering the minimum requirements or artefacts necessary to complete the analysis (Phelan, 2003). Without case knowledge the TCU analyst does not have a clear understanding of what is being looked for, so how does the analyst know where to look and when to stop looking (Pollitt, 2013). Consider photographs extracted from digital evidence related to a fraud investigation. A TCU analyst reviews the photographs and identifies those that are related to the fraud under investigation. As the sufficiency of evidence is met, the TCU analyst may not report on the hundreds of family and friend photographs extracted. A DFT member, when faced with reviewing the same photographs earlier in the process, notes that one of the photographs is of the suspect meeting a well known money launderer. This information would spawn an additional avenue of investigation that could possibly have been lost when the analysis is conducted away from the investigative team.

During training the DFT member is provided with guidance on things to consider in the absence of evidence. If no relevant artefacts are located the DFT member is trained to consider the possibility of encryption and assistance is provided by the TCU approved tool in identifying full disk encryption, or from installed programs. The DFT member should also consider the prioritization of the digital evidence and conduct a form of social analysis, effectively if the investigator and the DFT member feel that this item of digital evidence should contain the artefacts, then forward it to the parent TCU.

## Implementation

The first version of the DFT model was implemented six years ago by a parent TCU consisting of 25 members, of which 20 members were forensic analysts. The parent TCU was responsible for supporting approximately 8,500 employees policing Federal, Provincial, and Municipal regions covering 127 police stations ranging in size from two members to 800 members. The geographical area covered by the policing agency is approximately 945,000 square kilometres.

The DFT model was divided into two business lines:

- Digital Computer Field Triage (DCFT)
- Digital Mobile Filed Triage (DMFT)

*TCU approved tool*

An initial policy decision was to have the DFT members in the DCFT business line interrupt the boot process and conduct an assessment of the digital evidence in a forensically sound environment. With a large number of DFT members to be trained a commercially available tool was not viable due to the costs involved. Many of the tools, both commercial and open-source, were geared towards TCU analysts so were complicated to use or provided too much functionality for a non-digital evidence specialist. The decision was made to move towards a custom built Ubuntu boot disk, that would be designed, implemented and supported by the TCU. This TCU boot disk provided a simple text based menu for interaction with the DFT member, and supported searches designed around crime types.

One of the benefits of a custom built tool is that changes are able to be made from a grass roots level. An example of this was a request from a fraud investigator for credit cards numbers to be extracted, checked for validity against the Luhn algorithm and then sorted by bank code. This was implemented and the investigator was able to process digital evidence, extract possibly compromised credit card numbers and forward this information to the appropriate financial institutions. Reviews that previously took months, were being completed in hours, and the financial institutions were receiving information in time to prevent further financial losses.

As the DFT program expanded to deal with cell phones and other mobile devices, another review was conducted to determine the best tool for the DMFT program. The primary criteria for this decision was based on the tool that supported the majority of cell phones in the market. The decision at the time was to approve a commercial tool.

For both the DMFT and DCFT programs there does need to be a periodic review of the triage software and hardware available to ensure the best tool is being used in the DFT context.

*Training*

With the implementation of the Digital Field Triage model the selection of candidates is a critical first step. The candidates selected, and subsequently trained, will become ambassadors of the DFT model and the success of the implementation relies solely on these candidates. To ensure the appropriate candidate a pre-screening process is required to determine the level of computer skill the candidate possesses. This pre-screening is achieved through the use of a questionnaire for the candidate which covers both investigational and computer experience. A point to be considered is that the candidate is not applying for a forensic analyst position, so the level of computer skill need not be that high, however, a good working knowledge of computers is required.

With every course offering there are always more candidate applications than positions on the course, so a second pre-screening attribute is also used. This second attribute is based on the parent TCU needs, and covers the candidates geographical location and availability. Due to the large area covered consideration is given to selecting candidates that provide the maximum geographical coverage in an attempt to provide every investigator access to a DFT member. With multiple candidates applying from one police station consideration is given to the availability of the candidate and their assignment. This means that a candidate assigned as a detective, who is responsible for investigating serious crime and is not a first responder, would take priority over a candidate assigned to patrol, who is responsible to be a first responder out on the road.

*Digital Computer Field Triage course*

The DCFT course is five days in length and has a strong hands on style to provide the candidates with experience in handling digital evidence.

The first day provides the candidates with the expectations of their participation within the DFT program. Training is also provided covering TCU capabilities, handling digital evidence, legal considerations and investigational techniques. The candidates spend time setting up the laptop, installation and use of software, and set up of TCU approved tool. By reviewing how the candidates are able to complete these tasks does provide a further insight into the skill set of the candidate.

Day two provides training on how to interrupt the boot process of a computer and exposure to the capabilities of the TCU approved tool. The training details the steps needed to interrupt the boot process and to force the computer to use the TCU tool environment. For all the courses, the interrupting of the boot process was found to be one of the more stressful aspects of the course. The day continues with explanations of each of the searches the TCU approved tool provides. As each search is introduced the candidates reboot their own computers and practice the search on crafted evidence stored on a USB key. This is still an instructor led day, with candidates using what they have learnt. The final topic of the day is an explanation of the *Observation Report* and covers what is expected in the report.

Day three and four shift to the hands on component and the candidates work at their own pace. The candidates are provided with four different scenarios each on based on investigations they will be asked to assist on in the field. Each scenario has a dedicated platform, laptop computer, and a fully functional operating system with appropriate artefacts of the offence the candidates are investigating. The candidates are expected to follow all the policy and procedures they have been taught, and create an appropriate *Observation Report* with the additional files and information required. As each scenario is completed the reports are reviewed and appropriate feed back is given.

Day five is a testing day to determine if the candidates are capable of becoming DCFT members. In the morning the candidates are provided a test scenario which must be completed without the assistance of the instructors. As with the practice scenarios from days three and four, the candidates are faced with a scenario and digital evidence containing related artefacts. After the test scenario is completed a written test is given to the candidates covering the topics learnt on the course. Each of these tests are reviewed and marked by the instructors and only those candidates that have passed are certified as DCFT members.

*Digital Mobile Field Triage course*

The DMFT course is four days long, and again emphasis is placed on the hands on aspect of training.

As with the DCFT course day one is more lecture based, providing candidates with the same information related to the DFT program. For the DCFT course, training is provided on the commercial tool, however, the candidates are still trained in the setup, maintenance and use of the tool.

Days two and three consists class exercises with candidates being exposed to the primary platforms (iOS, Android, and BlackBerry) and conducting their own extractions for each device. Each candidate has a TCU approved tool for their primary use and instructors assist them with any problems. The third day does have more emphasis on the use of TCU approved tool and how it interacts with the previously extracted data.

Day four is an assessment day where the candidates are provide with scenarios for them to process individually and to answer a series of directed questions. Throughout the course instructors conduct assessments as to the suitability of the candidates to qualify as DMFT members.

### Continuing education

As with any part of the computer forensic field, techniques and digital evidence is constantly changing and there needs to be a way of passing this information onto the DFT members, both DCFT and DMFT. To facilitate this a forum has been set up on a dedicated server which only DFT members have access to. This forum provides the updated information and provides the ability to download instructional videos.

The other factor of ongoing training is that the skills learnt are perishable, and the DFT member's confidence deteriorates from lack of use. To maintain the ongoing use of the skills learnt there is a minimum number of assessments the DFT members must complete a year to maintain their qualification. In those areas where recent investigations have not produced digital evidence the DFT member is able to request a scenario from the parent TCU. This scenario is based on actual investigations and provides the DFT member with the ability to complete an assessment and follow the appropriate procedures.

### Management

The management of the program is integral to the success of the DFT model. Initially the management was shared between two senior analysts "off the side of their desk", but as the demand grew the support being received by the DFT members started to decline. Currently there is a dedicated forensic analyst who is responsible for the management of the DFT program.

The DFT coordinator is responsible for overseeing and managing all the DFT members, ensuring their level of competency in handling digital evidence meets the appropriate standard. On a practical level, the DFT coordinator will review all the *Observation Reports* created by the DFT members for both quality and accuracy, as well as identifying any training issues. For transparent management the DFT coordinator follows a defined policy that details what is expected of the DFT member. The policy covers the minimum number of calls a year, only TCU approved tools and equipment are to be used, and the ramifications of not following policy.

The role of DFT coordinator is important as the DFT members should never be working in isolation and require the oversight and support of a TCU. The DFT model is not a replacement for full analysis, but is part of the overall strategy of handling digital evidence. So an open line of communication between the TCU and DFT members is imperative. Along the same lines a DFT model is not to replace a TCU in smaller departments that do not have the need for a full time TCU. In cases such as this the smaller department still requires the oversight of a parent TCU, possibly from a different agency, but must follow the guidelines of the parent TCU.

To assist in the management of the DFT program a dedicated server has been set up for the DFT members. The same server used for continuing education is also the location for the storage of the most up to date version of the TCU approved tool. The server also provides links to every version of the tool released to DCFT members. The other function of the server is to provide the DFT members with a location to obtain a DFT file number, which is unique to their investigation. This file number is for the assessment of media by the DFT member on a particular investigation. If multiple DFT members work on the same file, then each DFT member is to obtain their own DFT file number. If there are multiple items of digital evidence to be assessed on one file, the DFT member would still only use the one DFT file number for their work on that file.

The DFT file number is recorded, along with the name of the DFT member who obtained it. This then provides the DFT coordinator with a metric of how many files each DFT member is conducting.

### Results

The primary objectives were:

1. Increase the efficiency of an investigation by providing artefacts from digital evidence in a timely manner.
2. Decrease the backlog of files for analysis by digital evidence specialists at a forensic laboratory.

One metric for the increased efficiency of an investigation is to record the number of DFT members available to investigators and are these DFT members being used. All the statistical information in the results section are derived from the implementation of the DFT program since 2009.

Table 1 shows the number of DFT members trained since 2009, and broken down by the business lines of the Digital Computer Field Triage and the Digital Mobile Field Triage. District 1 has the largest concentration of population and so the police stations responsible for this area are the largest, which is reflected in the highest concentration of DFT members. There is a consistent distribution of DFT members, providing good access for investigators. With this increased access comes the potential for increased efficiency in investigations.

**Table 1**
Locations of Digital Field Triage members across headquarters (HQ) and districts (D1–D4).

| DFT type | HQ | D1 | D2 | D3 | D4 | Total |
|---|---|---|---|---|---|---|
| DCFT members | 15 | 46 | 21 | 22 | 14 | 118 |
| - Police Stations Supported | 4 | 13 | 15 | 22 | 10 | 54 |
| DMFT members | 13 | 45 | 11 | 14 | 13 | 96 |
| - Police Stations Supported | 5 | 9 | 5 | 6 | 6 | 31 |

Since the inception of the DFT program there has been an increase in the number of investigations where the accused has entered a guilty plea in the time period between the DFT examination and exhibits being forwarded to the parent TCU. Discussions with investigators has determined that it is the review of the observation report by the Prosecutor and Defence Counsel that has partly led to the plea being entered. The comments received from the Prosecutor and Defence is the ability to see the artefacts related to the offence that assists in their decision.

Overall there is a general consensus by investigators that they are receiving actionable information from DFT members in a timely manner. The reviews of DFT files being submitted to the parent TCU does corroborate this consensus, however, this objective could be met solely on the fact that the investigators perception is that they are receiving the information when they need it and are not waiting for it.

The other objective was reducing the backlog faced by the parent TCU, and how this has been impacted by the use of the DFT model. The current backlog of the parent TCU is 58 files, of which 30 files were assessed by DFT members. Of the 30 files assessed by DFT members there is a reduction of exhibits being forwarded to TCU of approximately 75%, which means the forensic analysts are dealing with fewer items of digital evidence for each file. The reduction in time spent on a file, increases the number of files analysed per year, which in turn reduces the backlog.

Table 2 shows the correlation of an increase in DFT members constitutes a decrease in files sent to the parent TCU. As can be seen the three years prior to the introduction of the DFT model there was a steady increase in the number of files sent for TCU analysis. In 2009 the first DCFT course was taught, however, the levelling off of the files for analysis cannot be contributed to the DFT program. Within the parent TCU procedures were changed and a more selective process of files being accepted by TCU was put into place. Another reason is that the backlog of files was large and the turnaround time for analysis being completed was lengthy, thus there was a reluctance from investigators to send items of digital evidence to the TCU. These files not being sent to TCU were ones where digital evidence was a component of the investigation, but evidence from other areas secured the conviction. In 2011 DCFT training intensified in conjunction with increased awareness and access to the DFT program. From this we see a dramatic increase in the files being dealt with by DFT members, providing that immediate assistance to investigations. As more DFT members are trained in both business lines, the impact on files for TCU analysis can be seen.

With the number of TCU files for analysis starting to reduce, the backlog should also be reduced. The goal will never be to eradicate the backlog, but it is to bring it to a manageable level with an appropriate turnaround time.

*Review*

The ability to access trained DFT members has expanded the number of resources available to the parent TCU which is improving the efficiency of the entire program. During the implementation of the program there has been a shift in the location of the assessment of the digital evidence from the search scene to the office the DFT member works out of. In some investigations, DFT members attend the scene to assist with the identification and collection of the digital evidence only. The reason for the shift is so the DFT member is able to deal with the digital evidence in their own environment, with less of a perceived time pressure. There was also a problem of finding an area to conduct the assessment at scene due the physical state of the search scene. The DFT model is still being followed, just the location of the assessment has changed. The one drawback to this approach is the ability of the DFT member to identify attached USB devices, so assisting the search team in identifying other possible sources of digital evidence. A consideration here could be conducting a brief assessment of media at scene with the only purpose of identifying the attached devices and complete the remainder of the assessment at the local police station.

To ensure the integrity of the program a continuous assessment of all aspects needs to be conducted on a regular basis. The DFT coordinator position could conduct quality assurance by conducting a full analysis on a randomly selected investigation that has been assessed using the DFT model. This would then provide a better evaluation of the TCU tool and increase the credibility of the program. Equally the training course is evaluated after each of the offerings to maintain currency and provide the candidates with the best possible experience. It is important to build up the confidence of the candidates in a safe environment through training scenarios and the ability to give immediate feedback. The exam portions of the course are used to increase the stress of the candidates and an ability to review their knowledge, skills and ability. One change in the training is the make up of the instructor cadre which is now comprised of two TCU instructors and an experienced DFT member. The insight the DFT member provides in training is invaluable as it is coming from a person who is doing the tasks that the candidates will be expected to do. Another area identified for improvement is in candidate selection and determining a better metric to judge a candidate's ability.

**Table 2**
Digital Triage Files conducted.

| Year | Files | DFCT members | DMFT members | TCU files |
|---|---|---|---|---|
| 2006 | 0 | 0 | 0 | 345 |
| 2007 | 0 | 0 | 0 | 435 |
| 2008 | 0 | 0 | 0 | 526 |
| 2009 | 26 | 9 | 0 | 522 |
| 2010 | 73 | 0 | 0 | 480 |
| 2011 | 376 | 53 | 0 | 468 |
| 2012 | 265 | 81 | 0 | 476 |
| 2013 | 260 | 104 | 24 | 422 |
| 2014 | 409 | 118 | 84 | 329 |
| 2015 (June) | 469 | 118 | 96 | 137 |

The TCU approved tool for the DCFT program is working and meeting the desired goals, the problem it is facing is that it is custom built. This means that there is limited support and little business continuity. To avoid the costs of a commercial tool, which limits the number of DFT members trained from a budget perspective, there does need to be the inclusion of academia (James and Gladyshev, 2013). It is with the assistance of academia that adequate testing of tools can be achieved in a scientific manner and increase the trust and credibility of any tool. Law Enforcement investigators and forensic analysts have extensive knowledge of what is required to progress an investigation and secure a conviction. Academia is able to provide scientific support, recognised testing procedures and computer science specialists.

The final evaluation step is that the DFT program is being used and requested by the investigators it supports. There are constant requests for courses from supervisors who see the benefit of the DFT program. These requests are a metric for the programs success and acceptance by investigators. As the awareness and exposure of this DFT model expands, other law enforcement agencies are interested in training and setting up their own DFT program.

*Model evaluation*

The DFT Model was created following the principles of the Digital Science Research Process (Peffers et al., 2006) and so to be evaluated the following conditions should be addressed:

- Is it consistent with the models in the field of Digital Forensics?
- Is the model usable?
- Does the model guide the handling of digital evidence.

The basis of this model was the concepts identified in the Computer Forensics Field Triage Process Model (Rogers et al., 2006) which provided an outline for at scene triage. The DFT Model also does not replace laboratory procedures, for instance the methodology laid out in the United States Department of Justice Digital Forensic Analysis Methodology (2007). The integral part throughout is that the procedure described and used in the DFT Model is such that the forensic integrity of the digital evidence is maintained, and so any methodology following will receive digital evidence without accidental spoliation. So the DFT model is consistent with models already in the field of Digital Forensics.

A model is considered usable if it can be put into practice in a real life scenario and the desired outcome is achieved. To date 1878 DFT files have been created it which digital evidence was assessed and a determination was made as to whether further analysis was required by the parent TCU. While each one is not perfect, the model is being followed and understood, providing non-digital evidence specialists with a roadmap of dealing with digital evidence. The model also provides information on what courses of action to follow, and where there are inherent risks. This addresses the final two conditions that the model is usable and guides the handling of digital evidence.

**Conclusion and future work**

The focus of this paper was to offload some of tasks performed by forensic analysts to non-digital evidence specialists. The ISO 27037 (2012) does identify a role of DEFR, however, the job description for the DEFR is more than what a DFT member is allowed to do. The primary difference in the two roles is the ability to acquire the digital evidence. Consideration could be put to training a DFT member to complete the acquisition task, however, is this worthwhile. Within the corporate world, this could be beneficial, as the process is one of data collection, where the actual digital evidence often remains at the search scene. In a criminal investigation the digital evidence is part of the investigation and was potentially used for criminal activity. The majority of the time all the digital evidence is seized, under the authority of a search warrant, and removed from the search scene. The DFT member then determines which evidence would meet the threshold for further TCU examination, and the investigator is able to conduct a risk analysis as to whether or not to return those items that fail to meet the threshold. Additionally, there are circumstances where the digital evidence can never be returned, because it contains illicit material or information that would benefit the criminal enterprise. There is a definite role, however for a DFT member to acquire the contents of Random Access Memory (RAM) from a machine that is running. Over the last several years the importance of capturing RAM has increased along with the tools that can extract information from the RAM. The capturing of the RAM does require interaction with a live machine and so requires a greater confidence and ability in the DFT member. This leads to the concept of having advanced training for the DFT members.

The initial problem set was to create a Digital Field Triage (DFT) model to assist in increasing the investigational efficiency and reducing the backlog of digital evidence currently faced by Technological Crime Units. Through evaluation and implementation, it was found that the new DFT model met these requirements, with the increase in investigational efficiency being the most beneficial.

Investigators are receiving actionable information in a timely manner and fulfilling the need for intelligence led investigations, as opposed to reacting to information obtained months after the incident. It is this benefit to the investigator that shows the DFT program is an integral part of any handling of digital evidence process. In situations where there is a limited backlog there is still time taken in the process needed to send the digital evidence to the digital evidence specialist, or forensic analyst. Once received by the forensic analyst there is still the question of case knowledge and the "sufficiency of examination" both of which, potentially, could be of a detriment to the investigation. This is not to say that the DFT program should replace a TCU, as neither the DFT program nor TCU should operate in a vacuum. The DFT program relies upon a TCU to provide oversight and training, and the TCU relies on the DFT program to provide actionable information to the investigator. Due to the nature of the process further examination of digital evidence that meets the DFT program threshold is imperative, both for Court purposes, the need to have an expert testify, and that a more in-depth analysis will reveal further artefacts related to the offence. With a DFT program implemented the long term goal is to have every piece of digital evidence assessed prior to it being sent to a TCU. Forensic analysts at the TCU will then receive only items that are of investigational relevance and, as there are fewer items of digital evidence being forwarded, be able to spend more time on those items. If a forensic analyst is able to spend more time on an item of digital evidence, then it is possible to get past the "sufficiency of examination" (Phelan, 2003).

Within the digital forensic field there is still scepticism of the downloading of certain forensic tasks to non-digital evidence specialists (James and Gladyshev, 2013). The conclusions in that paper provide metrics to confirm that the scepticism is unfounded. More research in this area is needed, particularly in the reliability of the tools, but it is an area that must be pursued. In many jurisdictions law enforcement numbers are being reduced, whereas digital devices are becoming more and more prevalent. There is a need to expand the number of resources who have the knowledge, skills and aptitude to deal with digital devices. A tiered response mechanism provides this.

*Future work*

While the DFT model described as part of this paper has proved beneficial across numerous forensic investigations to date, there are a number of improvements possible to further improve the process including:

- Advanced training for experienced DFT members including:
  - RAM capture
  - Encrypted volume detection on live machine
  - Basic Acquisition
- Use of experienced DFT members as mentors/trainers.
- More accurate metrics to review the efficiency and effectiveness of the DFT model.

We also plan a comprehensive extension of this model to build a virtual platform for the purpose of training non-digital forensic investigators. This virtual platform will integrate different digital forensic scenarios such as observing and recording criminal activity, generating criminal intelligence, acquiring digital evidence (from network or local storage), live forensic tasks, etc. This platform could train non-digital forensic investigators skills required in analysis more complex systems, such as Industrial Control Systems (van Vliet et al., 2015), cloud computing platforms (Farina et al., 2015; Schut et al., 2015), etc.